# Recognizing Topic Change in Search Sessions of Digital Libraries based on Thesaurus and Classification System

Daniel Hienert and Dagmar Kern

GESIS – Leibniz Institute for the Social Sciences
Cologne, Germany
{firstname.lastname}@gesis.org

## ABSTRACT

Log analysis in Web search showed that user sessions often contain several different topics. This means sessions need to be segmented into parts which handle the same topic in order to give appropriate user support based on the topic, and not on a mixture of topics. Different methods have been proposed to segment a user session to different topics based on timeouts, lexical analysis, query similarity or external knowledge sources. In this paper, we study the problem in a digital library for the social sciences. We present a method based on a thesaurus and a classification system which are typical knowledge organization systems in digital libraries. Five experts evaluated our approach and rated it as good for the segmentation of search sessions into parts that treat the same topic.

## CCS CONCEPTS

• Information Systems~Users and interactive retrieval

## KEYWORDS

Digital Library; Search Session; Task, Topic Detection; Session Segmentation

## 1 Introduction

The analysis of transaction logs from real-world search engines shows that within a session, users do not only refine their queries over time but also handle different tasks and topics and that queries to these tasks are also interleaved [7]. The occurrence of different topics within one session makes it difficult to support the user over the whole session, e.g., with personalized search term recommendations and rankings.

In this paper, we study the problem of multiple topics and session segmentation in the world of digital libraries. In contrast to Web search, digital libraries often contain knowledge organization systems such as thesauri, classification systems or ontologies to organize their content. Documents in such digital libraries are explicitly tagged with keywords from a thesaurus and with categories from a classification system which can improve the overall retrieval effectiveness. In our approach, searched and viewed documents of a user session, their keywords and classifications are used to annotate user actions with topics. That forms the basis to segment a session into different topics. In the following, we present related work, followed by our approach, its evaluation and a discussion about the pros and cons.

## 2 Related Work

Gayo-Avello [3] provides an earlier survey on session detection methods. He defines search sessions as "short sequences of successive queries related to one single goal or information need" of a user. Several methods have been used to detect session boundaries based on time [e.g. 2], lexical analysis [e.g. 4], link and graph information [e.g. 1], categories and ontologies [8, 9], clustering and machine-learning approaches [e.g. 7] to combine different features.

So far, methods for topic detection in user sessions have been examined mostly for Web search. In contrast, digital library search often only applies time-based measures and rarely other approaches to segment sessions. For example, in [12] a sliding-window and a session-shift approach were used to segment PubMed user sessions. However, there are important differences between Web search and digital library or domain-specific search. In Web search, retrieved documents are Web pages, and queries can derive from all tasks and topics. In digital library search, documents are maintained by information professionals and are often organized by knowledge organization systems around a specific domain, community or topic. In this sense, this work is the first attempt to develop an algorithm for digital libraries which exploits the domain-specific thesaurus and classification system for the segmentation of user sessions to different topics.

## 3 Evaluation Environment

In this section, we introduce our testing environment. Sowiport [5] was a digital library for social science information with 9.7 million bibliographic records, full texts, and research projects. These come from 23 different databases with partly German and English focus. The portal reached about 25,000 unique visitors per week, mainly from German-speaking countries. Sowiport was discontinued at the end of 2017 in favor of the newly developed GESIS search[1].

By handling different databases in one search application, different search challenges arise. Each database uses a different thesaurus and classification system. For example, in the database for German Social Sciences Literature (SOLIS), documents are manually annotated by information professionals with keywords from the thesaurus for the social sciences (TheSoz)[2] and with categories from the classification for the social sciences[3]. The thesaurus contains about 12,000 entries with 8,000 descriptors and 4,000 synonyms. The classification system consists of 14 main classes and 145 subclasses. The other included databases use different knowledge systems which can lead to difficulties for users searching for a certain search term. That is why we implemented some services which reduce this effect and to which we refer later in this paper: (1) The heterogeneity service (HTS) contains cross-concordances for 25 different thesauri with 513,000 controlled terms [11]. It can be used, for example, to find equivalent terms in different thesauri. (2) The Search Term Recommender (STR) maps uncontrolled user search terms to thesaurus terms [10]. It is based on a co-occurrence analysis with free terms from titles and abstracts and controlled terms from the thesaurus.

## 4 Annotating Session Topics and Segmenting the Session

In the following, we show how a user search session can be annotated with keywords, categories and session topics. Based on that, the user session can be segmented. Figure 1 gives an overview of the process arranged on five different levels: (1) the user session with actions, (2) documents arising from these actions, (3) keywords from these documents, (4) resulting categories, and (5) session topics assigned to user actions. These levels will be explained in detail in the following section.

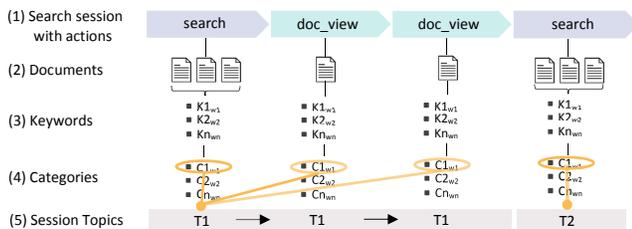

Figure 1: The process to annotate session topics.

### 4.1 Annotation Process

(1) The first level presents the user's *search process*, modeled as a sequence of user search queries ('search') and document views ('doc_view'). This means, the user conducts a query, receives a result list, and can inspect the result list for interesting documents. If a document seems to be interesting from the title, authors, source or snippet, the user can then inspect it in the detailed document view with a click on the title. There, she can e.g., read the abstract, bookmark it or read the full text.

[1] https://search.gesis.org/?lang=en
[2] http://lod.gesis.org/thesoz/en.html
[3] http://lod.gesis.org/thesozcl/en.html

(2) On the second level, both action types 'search' and 'doc_view' can be represented topically by a number of *documents*. A 'search' action, for example, can be represented by up to twenty results. A 'doc_view' action is simply represented by the document which has been viewed by the user.

(3) On the third level, each document from the level above can be represented by a number of *keywords* from the thesaurus for the social sciences (TheSoz, see Section 3). Documents with TheSoz terms can be taken directly. Documents with terms from other thesauri are transformed to TheSoz keywords with the HTS [11] service. If a document has no keyword information, we take the title, tokenize it, clean it from English and German stop words, take only words with more than three characters and transform them with the STR service [10] to TheSoz keywords. As a result, for each document, we have a topical representation by a list of keywords from the thesaurus. For a 'doc_view' action this keyword list can be taken directly to represent the action. To consider that documentalist for the SOLIS database push more specific and more important keywords at the beginning of the keyword list, we introduce a weighting factor ($w$ in Figure 1) for keywords. We use a discount formula to compute the weighting factor of the keyword in relation to its position $p$ in the keyword list: $Weight\_Keyword_p = 1/log_2(p + 1)$. For a 'search' action, we collect all keywords which appear in the first twenty documents of the result list. Keywords from the documents are weighted according to the discount formula above. But additionally, we take the position of the document in the result list into account. Here, we use a linear model to not give too much weight to the first positions, but rather have a smooth dumping factor over the whole list. For that, we use the formula $DocumentFactor_p = 1.05 - (0.05 * p)$ with $p = [1,20]$. This gives the first document in the list a factor of 1 and the tenth document in the list a factor of 0.55. The document factor is then used to additionally weight each keyword of a document. The final weighted keyword list can then be taken to represent the 'search' action.

(4) On the fourth level, the keyword lists are used to identify possible *categories* for the action from the classification system. For each keyword of the list, we query a look-up-table that contains a category for each keyword. This table is built based on the relationship of which keywords appeared most often with which category in the corpus of all SOLIS documents. By querying the list of keywords, we get a weighted list of categories which can again be used to describe the actions. Each user action of the search session is now described by documents, keywords, and categories. The keyword and the categories list both are ranked by the keywords appearing in the documents.

(5) On the fifth level, we want to add *session topics* to user actions that can reappear throughout the session. The goal is to have as few session topics as possible within a user session to recognize reappearing topics, but as much as necessary to recognize topic changes. In our approach, session topics are based on the categories of actions. Each action has a differently ranked list of categories. This can depend on the user search terms, chosen facets, the ranking of the search system and the document corpus. We, therefore, build a ranked list of categories by summing up all weights for a category over the whole session. This list is then used to re-sort the categories in each user action by pushing more common categories higher if the weighting of successive categories is close together. We are now able to select a session topic for every user action. For a 'search' action, we choose the top category from the action's category list (created on Level 4). For a 'doc_view' action, we choose the topic session from the previous 'search' action as it originates from the 'search' action.

## 4.2 Segmenting the Session

In the next step, we want to decide when a topic change appears in the session. Therefore, we want to add a *topic number* to each action. The algorithm utilizes two reasons for a reappearing session topic: (1) the session topic for two actions are the same. For example, two 'search' actions for 'facebook' and 'instagram' have both the session topic 'Interactive, electronic Media' from the classification system (remark: all examples in this paper come from original log data). Here, the search queries are not the same, but because of the classification category, we are able to merge them into one session topic. (2) The search queries have a search term in common. We use a Levenshtein distance of 2 to compare two terms from two different search queries. For example, search query 1 is 'migrant youth welfare sector' and search query 2 is 'migrants education'. These actions are then related because of the common term 'migrant'. The algorithm walks through the session from one action to the next. For each action, it then goes backward through the session and checks for the two rules by comparing the current action parameters with those from the session topics from the above actions. If it finds a similar session topic or search query, it takes that topic number, if not, it creates a new one. Figure 2 shows an example session with applied session topics, session numbers and the segmentation visualized as a red dashed line between the session segments.

| Action step | Action type | User search terms | Citation | Session topic | Topic number |
|---|---|---|---|---|---|
| [1] | Simple search | early childhood socialisation | | Social Psychology | T1 |
| [2] | Document View | | Neidhardt, Friedhelm (ed.) (1975): Early Childhood Socialisation: Theories and Analyses | Social Psychology | T1 |
| [3] | Document View | | Kirsch-Auwärter, Edit E. (1996): Gender Differences: Facts or Myths: A Plea for a New Look at Early Childhood Socialization | Social Psychology | T1 |
| [4] | Simple search | refugee policy | | Migration | T2 |
| [5] | Document View | | Angenendt, Steffen, et al. (2016) Many refugees, few data: fight-related development cooperation needs better data | Migration | T2 |

**Figure 2: Example session with two session topics (translated to English).**

## 5 Evaluation

In this evaluation, we want to understand the quality of the session topics generation and the quality of the session's segmentation for domain-specific digital library search. We would have liked to use an existing evaluation data set to make our results comparable. However, so far there exists only user session evaluation data for *Web search* (e.g. the AOL 2006 dataset or the TREC Session Track) that does not contain annotated documents with thesauri terms and categories from a classification system which our approach relies on. Therefore, we conducted an evaluation with five classification professionals who rated one hundred Sowiport user sessions.

### 5.1 Methodology

The evaluation of the quality of topic assignment and of the segmentation needs domain experts, but also experts in the classification system. Evaluators need to be able to assess if a session topic fits the user's search or document view action. This requires on the one hand knowledge about the topics social sciences users are searching and looking for and on the other hand knowledge about the systematics of the classification system. We asked five classification experts to rate the quality of the session topic assignment and the quality of the session segmentation. Each expert works or had worked daily with the classification system and has several years of experience.

We build an online tool with which the experts can easily assess the quality of session topics and segmentation. The tool shows a user session as a list of user actions (cp. Fig. 2). For a simple or advanced search, it shows the user search terms, for a faceted search the clicked facet terms. For a user's document view, we show the document's citation in APA style (favored in the social sciences). For each user action, the computed session topic is shown. The session segmentation is shown in the column 'topic number' and additionally with red lines in the table between the different segments.

From the Sowiport transaction logs, we have built a one-year dataset of user sessions from 01/08/2016 to 31/07/2017. Therefrom, we automatically selected 100 sessions with 2 to 30 user actions and a maximum session duration of two hours. For each number of user actions up to four different sessions were taken to consider different user activity levels. The resulting evaluation dataset contained 100 sessions with 1,145 actions (11.45 on average, min: 2, max: 26). For different actions, we have 567 document views, 489 simple searches, 64 facet searches, and 25 advanced searches. A session lasts on average for about 26 minutes.

Assessors can then rate both (1) the quality of the session topic assignment and (2) the quality of segmentation on a five-point Likert scale for quality with the ratings "very bad"=-2, "bad"=-1, "acceptable"=0, "good"=1, "very good"=2 and optionally "do not know". Additionally, for each session, a comment can be given. An assessor can click through the 100 sessions, one by one, and can assess each. Experts can take a break anytime and can continue the process later on.

### 5.2 Results

The quality of the session topic assignment was rated on average with 0.279 ("acceptable"). The quality of session segmentation was rated on average with 0.833 ("good"). The interrater agreement measured with the Intraclass Correlation Coefficient (ICC) was 0.825 for the assignment and 0.622 for the segmentation part. After the evaluation, we conducted short interviews concerning the shortcomings and pitfalls of our approach. The assessors' main concern was that user queries or documents with more than one topic were mapped to only one single session topic. This is difficult in their point of view, as in their daily workflow, they are assigning several keywords and one or more classification to a single document.

# 6 Discussion

Recognizing the task topic only from log data is challenging. Especially in digital library search, where search queries can be complex and specialized for the domain. The input data for such an approach are (1) the user queries themselves and (2) the user's interaction with the system. Related work has built on simpler features of user interaction (such as only time) or more complex ones (such as combinations of query, session, history, clicks, dwell times). However, it is purposeful to add additional knowledge sources to the input data to find (a) broader categories of task topics and (b) semantically related queries that belong together. Liu et al. [8] used domain-independent categories from ODP. Hua et al. [6] and Lucchese at al. [9] mapped search queries to outer knowledge concepts to find relations between related queries.

In our approach, we exploit domain knowledge by enriching each user interaction with keyword distributions from the thesaurus and category distributions from the classification system, the basis for session topics. This reduces complexity as uncontrolled user language from search queries is mapped to the controlled language of thesaurus and classification. Thereby, the approach showed some advantages, especially for search interactions which are typical for digital libraries: (a) the assignment of keywords, categories and session topics to user actions gives the algorithm more knowledge ground than by only comparing queries lexically or semantically. For example, semantically related terms such as "climate change" and "greenhouse effect" are both mapped to the session topic "Ecology, Environment". "Instagram" and "Pinterest" are mapped to "Interactive, electronic Media". (b) Also, searches for authors (which are common in domain-specific search) can be mapped to a session topic if the author is specialized in one. For example, a query for "Bernhard Nauck" is mapped to "Family Sociology", "Walther Specht" is mapped to "Social Work". The same applies to other kinds of typed searches such as for journals or proceedings. (c) Uncontrolled user terms and complex term combinations are also mapped to a session topic. (d) If a user session only consists of document views, e.g., by browsing over related documents, a session topic is found.

There are also disadvantages if this approach is used alone. The quality depends strongly on the different parts of the system, namely the thesaurus and classification system, the documents corpus, the quality of tagging documents and on the retrieval system. This can result in some issues: (1) Too broad user queries or result lists containing many documents with broad classifications could lead to very broad session topics such as "General Sociology". (2) The session topic can switch between mostly similar topics in the same session, just because one part is tagged with the top category, the other with the subcategory. (3) A user query or a document with several topics is mapped to a session topic that is dominant in the overall session.

The human experts rated the quality of session topic assignment on a mid-range. The main reason is that user actions are mapped to only one single session topic although the session contains most often multiple topics. Additionally, session topics are chosen from the category list in a way that as few session topics as possible are selected for the whole session. Choosing only a few session topics is beneficial for the session segmentation because it guarantees a better clustering of user actions with similar topics.

The approach of session segmentation is based on two features: session topics and query content. Session topics were computed directly from the document or from the search results in real-time making the approach applicable also in a live environment. With these simple features, we are able to achieve a good quality of segmentation. Combining these basic features with additional ones (lexical, semantical, temporal, session) could lead to even better results.

# 7 Conclusion

In this work, we proposed a new method for the annotation of session topics and the segmentation of a whole user session. The method is based on typical knowledge organization systems in digital libraries such as thesauri and classification systems. The approach showed some advantages for digital library search as an addition to existing features: (1) Semantically related query terms fall under the same session topic, (2) author searches fall into the session topic of their expertise, (3) proceedings and journal searches get a session topic for their subject, (4) free user search terms not contained in a thesaurus get a session topic, (5) searches with multiple terms get a unique session topic, and (6) sessions with only document views get one or more session topics. Five independent expert evaluators rated the new method as good for the topical segmentation of search sessions.